\documentclass[aps,preprint,a4paper,amsmath,amssymb,floatfix]{revtex4}
\usepackage{accents}
\usepackage{graphicx}

\newcommand{\ket}[1]{\ensuremath{\left|{#1}\right \rangle}}
\newcommand{\bra}[1]{\ensuremath{\left\langle{#1}\right|}}
\newcommand{\braket}[2]{\ensuremath{\left\langle{#1} \left|\right.{#2} \right\rangle}}

\newcommand{\op}[2]{\ensuremath{{\bf \hat {#1}} {#2} }}
\newcommand{\com}[1]{\ensuremath{\left[ {#1} \right]}}
\DeclareMathOperator{\tr}{Tr}

\newlength{\dhatheight}
\newcommand{\supopDS}[1]{%
    \settoheight{\dhatheight}{\ensuremath{\hat{#1}}}%
    \addtolength{\dhatheight}{-0.35ex}%
    \hat{\vphantom{\rule{1pt}{\dhatheight}}%
    \smash{\hat{#1}}}}
\newcommand{\supopTS}[1]{%
    \settoheight{\dhatheight}{\ensuremath{\hat{#1}}}%
    \addtolength{\dhatheight}{-0.35ex}%
    \hat{\vphantom{\rule{1pt}{\dhatheight}}%
    \smash{\hat{#1}}}}
\newcommand{\supopS}[1]{%
    \settoheight{\dhatheight}{\ensuremath{\scriptstyle{\hat{#1}}}}%
    \addtolength{\dhatheight}{-0.175ex}%
    \hat{\vphantom{\rule{1pt}{\dhatheight}}%
    \smash{\hat{#1}}}}
\newcommand{\supopSS}[1]{%
    \settoheight{\dhatheight}{\ensuremath{\scriptscriptstyle{\hat{#1}}}}%
    \addtolength{\dhatheight}{-0.07ex}%
    \hat{\vphantom{\rule{1pt}{\dhatheight}}%
    \smash{\hat{#1}}}}
\newcommand{\hhat}[1]{\ensuremath{ \mathchoice{\supopDS{#1}}{\supopTS{#1}}{\supopS{#1}}{\supopSS{#1}}}}
\newcommand{\supop}[1]{\hhat{\mathcal{#1}}}

\begin{document}

\title{The Scaling of Weak Field Phase-Only Control in Markovian Dynamics}
\author{Morag Am-Shallem and Ronnie Kosloff}
\affiliation{Fritz Haber Research Center and the Institute of Chemistry, the Hebrew University, Jerusalem 91904, Israel}

\begin{abstract}
We consider population transfer in open quantum systems, which are described by quantum dynamical semigroups (QDS).
Using second order perturbation theory of the Lindblad equation,
we show that it depends on a weak external field only through the field's autocorrelation function (ACF), which is phase independent. 
Therefore, for leading order in perturbation, QDS cannot support dependence of the population transfer on the phase properties of weak fields.
We examine an example of weak-field phase-dependent population transfer, 
and show that the phase-dependence comes from the next order in the perturbation.
\end{abstract}

\maketitle

\section{Introduction}
\label{sec:intro}
Quantum control is devoted to steering a quantum system toward a desired objective.
Coherent control achieves this goal by manipulating interfering pathways via external fields, 
typically  a shaped light field \cite{rice1992new}. 
Early in the development of quantum control,
Brumer and Shapiro  proved that for weak fields in an isolated system phase only, control is impossible 
for an objective which commutes with the free Hamiltonian \cite{brumer_one-photon-control_1989}.
A qualitative explanation is that under such conditions there are no interfering pathways 
leading from the initial to the final stationary states.

More formally, the control electromagnetic field  in the time domain is  $\epsilon(t)$, 
and its spectrum is given by 
\begin{equation}
\label{eq:amp_and_phase}
\tilde{\epsilon}(\omega) = \tilde{A}(\omega) e^{i\tilde{\varphi}(\omega)},
\end{equation}
where $\tilde{A}(\omega)$ is the amplitude and $\tilde{\varphi}(\omega)$ is the phase. 
Any target operator that commutes with
the field independent Hamiltonian \op{H}{} is uncontrollable by 
the phase  $\tilde{\varphi}(\omega)$ \cite{spanner_one-photon-control_2010}.

Experimental evidence has challenged this assertion. 
Weak field phase-only (WFPO) control was demonstrated first by Prokhorenko et al. 
\cite{prokhorenko_population-transfer_2005,prokhorenko_retinal-control_2006}. 
The target of control was an excited state branching ratio.
The phenomena was attributed to the influence of the environment. 
A subsequent study by van der Walle et al. showed that such controllability is solvent dependent 
\cite{vanderwalle_solvation-induced_2009}. 

A careful examination of the assumptions can resolve
the discrepancy between theory and experiment, considering 
that the experiments were carried out for an open quantum system.
It has been suggested that the coupling to the environment changes the conditions 
under which the statement of impossibility holds. A new relaxation timescale emerges which interferes with
the timescale influence by the pulses phase.
Numerical evidence that WFPO control becomes possible for an  open quantum system was shown by Katz et al.  
\cite{katz_weak-field-control_2010}. In line with the original proof, 
Spanner et. al. \cite{spanner_one-photon-control_2010}
argued that if the coupling between the system and the environment does not commute with the measured observable,
then the conditions for phase insensitivity do not hold.
Nevertheless, open quantum systems have additional features which are not covered by the  Hamiltonian time dependent
perturbation theory employed to prove the WFPO no go result. A possible opportunity for WFPO for control of
observables commuting with the Hamiltonian can emerge from the continuous nature
of the spectrum of the evolution operator and or the inability to separate the system from its environment.

To clarify this issue we will explore the conditions which enable or disable WFPO control in an open quantum system.
We restrict this study to the axiomatic approach of open systems  based on quantum dynamical semigroups (QDS).
The theory aims to find the propagator of the reduced dynamics of the primary system 
under the assumption that it is generated by a larger system bath Hamiltonian scenario. The generator in this case
belongs to the class of completely positive maps \cite{kraus1971}. 
An important consequence is that the system and bath
are initially uncorrelated or, formally, are in a tensor product state at $t=0$. 
An additional assumption is the Markovian dynamics. 
Under completely positive conditions, Lindblad and Gorini-Kossakowski-Sudarshan (L-GKS) proved that the Markovian 
generator of the dynamics \supop{L} has a unique structure \cite{lindblad_generators_1976,gorini_QDS_1976}.  
This generator extends the system-bath separability assumption to all times. 
The WFPO controllability issue can be related now to observables which are invariant to the field free dynamics.

To shed light on the existence/nonexistence of weak field phase only control for L-GKS dynamics
we examine the control of population transfer which is an invariant of the field free dynamics.
The population transfer $\Delta N$ can be directly observed experimentally for fluorescent dyes with  a unit quantum yield.
A complementary experiment is the weak field spectrum of a photo absorber in solution.
For both types of experiments WFPO control of population
will lead to phase sensitivity of weak field spectroscopy.
 
The main result of the present study is that population transfer and energy absorption spectroscopy in L-GKS dynamics depends, in the  leading order, 
only on the  autocorrelation function (ACF) of the field, defined by 
\begin{equation}
\label{eq:ACF_definition}
C(\tau) = \int\limits_{-\infty}^{\infty} \,dt \epsilon(t+\tau)\epsilon^*(t).
\end{equation}
The ACF does not depend on the phase of the field $\tilde{\varphi}(\omega)$ (cf. Appendix \ref{sec:app_phase_indep}). 
Therefore phase-dependent control of population transfer will take place only in the next order of the field strength.

\section{The model}

Consider a molecule with two potential electronic surfaces, \op{H}{_g} and \op{H}{_e}, 
coupled with a weak laser field $\epsilon(t)$ through the field operator $\op{V}{(t)}$.
Starting with an initial state in the ground electronic surface \ket{\psi_0}, 
the control objective is the population transfer to the excited surface.
The system Hamiltonian and the field operators are, respectively:
\begin{equation}
\begin{array}{cc}
\op{H}{_0} = \left( \begin{array}{cc} \op{H}{_e} & 0 \\ 0 & \op{H}{_g} \\ \end{array} \right),
&
\op{V}{(t)} = \left( \begin{array}{cc} 0 & \op{\mu}{} \epsilon(t)  \\ \op{\mu}{} \epsilon(t)^{*} & 0 \\ \end{array} \right),
\\
\end{array}.
\end{equation}

The control objective is the projection on the excited electronic surface:
\begin{equation}
\op{P}{_e} = \left( \begin{array}{cc} \op{1}{_e} & 0 \\ 0 & 0 \\ \end{array} \right).
\end{equation}
This objective commutes with the field free Hamiltonian $[\op{P}{_e},\op{H}{_0} ]=0$.

The population transfer is calculated  by solving for the dynamics of the density operator $\op{\rho}{}$  in Liouville space.
The L-GKS equation generates the dynamics:
\begin{equation}
i \hbar \frac{\partial \op{\rho}{}}{\partial t} = \supop{L} \op{\rho}{},
\end{equation}
Where $ \supop{L} =\supop{L}_0 +\supop{V}(t) $ is the Lindbladian, 
and $\op{\rho}{} = \sum \rho_{cd}\ket{c}\bra{d}$ is the density operator
represented here in the Hamiltonian eigenstates basis, 
$ \op{H}{_0}\ket{c} = \hbar \omega_c \ket{c}. $
The action of the superoperator $\supop{V}$ on the density matrix $\op{\rho}{}$ is defined by:
\begin{equation}
\supop{V} \op{\rho}{} = \com{\op{V}{},\op{\rho}{}}.
\end{equation}
For a specific element of the density operator \ket{c}\bra{d} it yields:
\begin{equation} \label{eq:commutator_operation}
\supop{V}(t) \ket{c}\bra{d} = 
\sum\limits_m \left(\epsilon(t)\mu_{mc}\ket{m}\bra{d} - \epsilon^*(t)\mu_{dm}\ket{c}\bra{m}\right).
\end{equation}
The action of $\supop{L}_0$ is more involved. 
Under the complete positivity and the Markovian assumptions,
the general L-GKS expression is \cite{lindblad_generators_1976,gorini_QDS_1976}
\begin{equation}
\supop{L}_0 \op{\rho}{} = \com{\op{H}{_0},\op{\rho}{}} 
+ i \sum\limits_k \left( \op{A}{_k} \op{\rho}{} \op{A}{_k^\dagger} 
-\frac{1}{2}\left( \op{A}{_k^\dagger} \op{A}{_k} \op{\rho}{} + \op{\rho}{} \op{A}{_k^\dagger} \op{A}{_k} \right)\right).
\end{equation}
Where $\op{A}{}$ is an operator defined in the systems Hilbert space.
The commutator with the Hamiltonian governs the unitary part of the dynamics,
while the second term on the rhs leads to dissipation and dephasing.
Notice that the target operator is invariant to the dissipative dynamics $\supop{L}_0^* \op{P}{_e}=0$.

The initial state is an equilibrium distribution $P(a)$ on the ground electronic surface:
\begin{equation} \label{eq:init_rho}
\op{\rho}{_0}=\sum\limits_{a\in g.s.}P(a)\ket{a}\bra{a}.
\end{equation}

The population transfer in this case is $\Delta \op{N}{}= \op{P}{_e}$. $\op{P}{_e}$ will be calculated by means of second order time dependent perturbation theory of L-GKS equation.
This is the lowest order that yields population transfer.
In the case of unitary dynamics, i.e. $ \supop{L}_0 \op{\rho}{} = \com{\op{H}{_0},\op{\rho}{}} $,
it yields the same results as the equivalent calculation by the first order perturbation theory of the Schr\"odinger equation.
In the same manner, the next order in population transfer calculation is the the fourth power of the field strength.

\section{Population transfer in Liouville space}

The lowest order of population transfer, starting from the initial condition of Eq. (\ref{eq:init_rho}),
is calculated employing second order time dependent perturbation theory: 
\begin{equation}
\Delta N(t) 
= \left\langle \op{P}{_e} \right\rangle(t) 
= \tr\left\lbrace \op{P}{_e} \op{\rho}{_I} \right\rbrace
\approx \tr\left\lbrace \op{P}{_e} \op{\rho}{_I}^{(2)}(t) \right\rbrace
\end{equation}
where $\op{\rho}{_I}(t_f)$ is the density matrix in the interaction picture, at the final time $t_f$,
and 
\begin{equation}
\op{\rho}{_I}^{(2)}(t_f) = 
\left(-\frac{i}{\hbar}\right)^2 \int\limits_{t_i}^{t_f} \,dt_2 \int\limits_{t_i}^{t_2} \,dt_1 
e^{-i \supop{L}_0 (t_f-t_2)} \supop{V}(t_2) 
e^{-i \supop{L}_0 (t_2-t_1)} \supop{V}(t_1)  
e^{-i \supop{L}_0 (t_1-t_i)} \op{\rho}{_0}
\end{equation}
is the second order perturbation term in the interaction picture.

Before we evaluate this expression in some representative cases, it can be simplified.
First, we note that if initially the system is in equilibrium and invariant to $\supop{L}_0$ , then
\begin{equation}
e^{-\frac{i}{\hbar} \supop{L}_0 (t_1-t_i)} \op{\rho}{_0} = \op{\rho}{_0}.
\end{equation}  
Next, the order of the left operations can be changed leading to:
\begin{equation}
\tr\left\lbrace \op{P}{_e} \int\limits_{t_i}^{t_f} \,dt_2 \int\limits_{t_i}^{t_2} \,dt_1 
e^{-\frac{i}{\hbar} \supop{L}_0 (t_f-t_2)} \op{\rho}{} \right\rbrace
= \int\limits_{t_i}^{t_f} \,dt_2 \int\limits_{t_i}^{t_2} \,dt_1  \tr\left\lbrace 
e^{-\frac{i}{\hbar} \supop{L}_0 (t_f-t_2)} \op{P}{_e} \op{\rho}{} \right\rbrace,
\end{equation}
and, since Lindbladian dynamics preserves the trace then:
\begin{equation}
\tr\left\lbrace e^{-\frac{i}{\hbar} \supop{L}_0 t_2} \op{P}{_e} \op{\rho}{} \right\rbrace = \tr\left\lbrace \op{P}{_e} \op{\rho}{} \right\rbrace,
\end{equation}
it yields:
\begin{equation} \label{eq:pop_trans_expr}
\Delta N(t_f) = -\frac{1}{\hbar ^2}   
\int\limits_{t_i}^{t_f} \,dt_2 \int\limits_{t_i}^{t_2} \,dt_1 
\tr\left\lbrace \op{P}{_e}
\supop{V}(t_2) e^{-\frac{i}{\hbar} \supop{L}_0 (t_2-t_1)}\supop{V}(t_1)  \op{\rho}{_0}
\right\rbrace.
\end{equation}

Eq. (\ref{eq:pop_trans_expr}) is now evaluated  in unitary and non-unitary dynamics. 
See Appendix \ref{sec:app_calculations} for detailed calculations.

\subsection{Unitary dynamics generated by the Hamiltonian}

In this case we get:
\begin{equation} 
\label{eq:integral_unitary}
\Delta N = \sum\limits_{\substack {a\in g.s. \\ b\in e.s}} P(a) 
\frac{\left| \mu_{ab} \right|^2}{\hbar^2}
\left\lbrace \int\limits_{0}^{\infty} \,d\tau 
C^*(\tau) e^{-i \omega_{ba}\tau} + c.c. \right\rbrace.
\end{equation}
where $\mu_{ab}$ is a matrix element of the operator $\op{\mu}{}$ in the energy basis, 
$C^*(\tau)$ is the complex conjugate of the autocorrelation function (ACF) of the field $\epsilon(t)$, 
defined in Eq. (\ref{eq:ACF_definition}), 
and $ \omega_{cd}\equiv\omega_{c}-\omega_{d} $.
$g.s.$ and $e.s.$ denote the ground and excited surfaces, respectively.
$c.c.$ denotes the complex conjugate.

The ACF does not depend on the phase of the field. 
This is shown in Appendix \ref{sec:app_phase_indep} by means of the vanishing of the functional derivative of the ACF with respect to the phase.
Therefore, the population transfer is not affected, to this order in the field strength, 
by the phase properties of the field.
This result is not new \cite{spanner_one-photon-control_2010}.
It is presented here in order to demonstrate the perturbative calculation in Liouville space
and to emphasize the dependence on the ACF.

\subsection{General L-GKS dynamics}
	
In the present study, the L-GKS generator can only induce dephasing and relaxation within the electronic surfaces.
Electronic dephasing or electronic relaxation for which $\op{P}{_e}$ is  invariant are  considered.
As a result, population transfer is generated only by \supop{V} (the commutator of \op{V}{}).

The notation is simplified using the fact that all states in the perturbation expansion are filtered by  \op{\mu}{}.
We define $\ket{\theta_a}\equiv \op{\mu}{}\ket{a}$  (or $\bra{\theta_a}\equiv \bra{a}\op{\mu}{}$, respectively), 
and it should be understood as a state projected on the excited electronic surface. 
We will also use the notation $\op{\Theta}{_a} \equiv \ket{\theta_a}\bra{a}$ for the relevant density matrix element. 
With this notation, the expression in Eq. (\ref{eq:pop_trans_expr}) becomes:
\begin{equation}
\label{eq:integral_LGKS}
\Delta N = \frac{1}{\hbar ^2}
\sum\limits_{\substack {a\in g.s. \\ b\in e.s}} P(a) 
\int\limits_{0}^{\infty} \,d\tau \left( C^*(\tau)
\bra{b}\left[ e^{-\frac{i}{\hbar} \supop{L}_0 \tau} \op{\Theta}{_a}\right]\ket{\theta_b} 
+c.c. \right)
\end{equation}
To proceed beyond this point additional details on the operation of \op{\mu}{} and $\supop{L}_0$ 
are required. 
Nevertheless the dependence on the control field is only through its ACF.

\subsection{General non unitary dynamics}

These results can be extended to a more general propagator $\supop{U}_0(t_f,t_i)$.
The conditions are:
\begin{enumerate}
\label{sec:4axioms}
	\item The dynamics under weak fields can be described by a second order perturbation theory:
\begin{equation}
\op{\rho}{_I}^{(2)}(t_f) = 
\left(-\frac{i}{\hbar}\right)^2 \int\limits_{t_i}^{t_f} \,dt_2 \int\limits_{t_i}^{t_2} \,dt_1 
\supop{U}_0 (t_f,t_2) \supop{V}(t_2) 
\supop{U}_0 (t_2,t_1) \supop{V}(t_1)  
\supop{U}_0 (t_1,t_i) \op{\rho}{_0}
\end{equation}
	\item The field-free propagation is homogeneous in time, and therefore depends only on the time difference:
\begin{equation} \label{eq:homogeneousEvolution}
\supop{U}_0(t_b,t_a) = \supop{U}_0(t_b-t_a)
\end{equation}
	for any $t_a,\ t_b$
	\item The initial density matrix is invariant under the field-free propagator:
\begin{equation}
\supop{U}_0(t) \rho_0 = \rho_0
\end{equation}
	\item The field-free propagator does not couple the two electronic surfaces: 
\begin{equation}
\tr\left\lbrace \supop{U}_0 (t) \op{P}{_e} \op{\rho}{} \right\rbrace = \tr\left\lbrace \op{P}{_e} \op{\rho}{} \right\rbrace.
\end{equation}
\end{enumerate}

Under these conditions, we can get the ACF-dependent expression:
\begin{equation}
\label{eq:integral_general_non_unitary}
\Delta N = \frac{1}{\hbar ^2}
\sum\limits_{\substack {a\in g.s. \\ b\in e.s}} P(a) 
\int\limits_{0}^{\infty} \,d\tau \left( C^*(\tau)
\bra{b}\left[ \supop{U}_0 (\tau) \op{\Theta}{_a}\right]\ket{\theta_b} 
+c.c. \right)
\end{equation}

\section{The relation between population transfer and energy absorption}
\label{sec:pop-energy}

Spectroscopy is based on using a weak probe to unravel pure molecular properties. 
Absorption spectroscopy measures the energy absorption from the field. 
Here, we relate 
this quantity to  the population transfer measured by delayed fluorescence.
We show that in a weak field under the L-GKS conditions also the energy absorption is independent  of the phase of the field.
In the adiabatic limit, i.e., for a slowly varying envelope function, 
this relation can be deduced directly from the expression for the population transfer.
For the  non-adiabatic cases, we prove an additional theorem.

\subsection{Adiabatic limit}
The power absorption is derived from the Heisenberg equation of motion:
\begin{equation}
\mathcal{P} = \frac{d \left\langle E \right\rangle}{dt} 
= \left\langle \frac{d \op{H}{}}{dt} \right\rangle
= \left\langle \frac{d \op{V}{(t)}}{dt} \right\rangle
= \left\langle  \left( 
\begin{array}{cc} 
0 & \op{\mu}{} \frac{\partial \epsilon}{\partial t}  \\ 
\op{\mu}{} \frac{\partial \epsilon^{*}}{\partial t} & 0 \\ 
\end{array} 
\right) \right\rangle,
\end{equation}
The expectation value of an operator $\op{A}{}$ is defined as
$\left\langle \op{A}{} \right\rangle = \text{tr}\left(\op{A}{} \op{\rho}{}\right)$.
We separate the density matrix to the populations on the upper and lower electronic surfaces $\op{\rho}{_e}$, $\op{\rho}{_g}$,
and for coherences $\op{\rho}{_c}$, $\op{\rho}{_c^\dagger}$:
\begin{equation}
\op{\rho}{} = 
\left( 
\begin{array}{cc} 
\op{\rho}{_e} & \op{\rho}{_c}  \\ 
\op{\rho}{_c^\dagger} & \op{\rho}{_g} \\ 
\end{array} 
\right),
\end{equation}
leading to the power absorption: 
\begin{equation}
\mathcal{P} = 
\text{tr}\left(
\frac{\partial \epsilon}{\partial t}\op{\mu}{} \op{\rho}{_c}
+ \frac{\partial \epsilon^{*}}{\partial t}\op{\mu}{} \op{\rho}{_c^\dagger}
\right)
= 2 \mathcal{R}e \left( 
\frac{\partial \epsilon}{\partial t}
\text{tr} \left( \op{\mu}{} \op{\rho}{_c} \right) \right).
\end{equation}

Total energy absorption is obtained by integrating the power:
\begin{equation}
\label{eq:deltaE_exact}
\Delta E(t_f) = 2 \mathcal{R}e \int\limits_{t_i}^{t_f}   
\frac{\partial \epsilon}{\partial t}
\text{tr} \left( \op{\mu}{} \op{\rho}{_c(t)} \right) \,dt.
\end{equation}

Similarly, the total population transfer is given by:
\begin{equation}
\label{eq:deltaN_exact}
\Delta N(t_f) = -\frac{2}{\hbar} \mathcal{I}m \int\limits_{t_i}^{t_f}   
\epsilon(t) \text{tr} \left( \op{\mu}{} \op{\rho}{_c(t)} \right) \,dt.
\end{equation}

The changes in energy and population are related. 
If we factorize the field to an envelope $\Lambda(t)$ and fast oscillations with the carrier frequency $\omega_L$:
\begin{equation}
\epsilon(t) = \Lambda(t) e^{i \omega_L t},
\end{equation}

then we can write:
\begin{equation}
\Delta E(t_f) = 2 \mathcal{R}e \int\limits_{t_i}^{t_f}   
\left( 
i \omega_L \epsilon(t) 
+ \frac{\partial \Lambda}{\partial t} e^{i \omega_L t}
\right)
\text{tr} \left( \op{\mu}{} \op{\rho}{_c(t)} \right) \,dt.
\label{eq:detf}
\end{equation}

In the adiabatic limit, i.e. for a slowly varying envelope function, i.e. 
\begin{equation}
\label{eq:adiabatic_condition}
\frac{\partial \Lambda / \partial t}{\Lambda} \ll \omega_L , 
\end{equation}
the second term is negligible.
Then we can write \cite{kosloff1992excitation,ashkenazi1997quantum}:
\begin{equation}
\label{eq:deltaE_adiabatic}
\Delta E \approx \hbar \omega \Delta N.
\end{equation}

In such cases we use the expressions derived above for the population transfer 
(Eqs. (\ref{eq:integral_unitary}), (\ref{eq:integral_LGKS}) and (\ref{eq:integral_general_non_unitary})) 
to obtain the phase independence of the energy spectrum.

\subsection{Non-adiabatic treatment}

In the nonadiabatic case, we have to evaluate the second term in Eq. (\ref{eq:detf})  using  the second order perturbation theory.

The coherence $\op{\rho}{_c(t)}$ is evaluated  from the first order expression for the density matrix:
\begin{equation}
\op{\rho}{_I^{(1)}} (t) = 
-\frac{i}{\hbar} \int\limits_{t_i}^{t} \,dt_1 
e^{i \supop{L}_0 t_1} \supop{V}(t_1)  
e^{-i \supop{L}_0 (t_1-t_i)} \op{\rho}{_0}
\end{equation}
Next, we substitute $\op{\rho}{_I^{(1)}} (t)$ in  the expression for  energy absorption, 
Eq. (\ref{eq:deltaE_exact}),  integrate and manipulate as described  in Appendix \ref{sec:app_calculations}. 
The result is that the energy absorption has a functional dependence on 
the cross-correlation function of the field with its derivative:
\begin{equation}
\int\limits_{-\infty}^{\infty} \,dt \epsilon(t)
\left.\frac{\partial\epsilon^*}{\partial t}\right|_{\tau+t}.
\end{equation}

However, this expression is also phase-independent. 
This can be shown using the functional derivative with respect to the phase of the field,
cf. Appendix \ref{sec:app_phase_indep}.

\section{Discussion}
We demonstrated  that, in general, 
the weak-field spectroscopy is functionally dependent 
only on the auto correlation function of the field. 
As a result, phase sensitivity is absent. 
This remains true even when the dynamics is generated by the Markovian
L-GKS equation. 
Moreover, 
this is also true for non-Markovian dynamics, generated by the time independent Hierarchical Equations of Motion approach (HEOM) 
\cite{yan_HEOM_2008,tanimura_HEOM_2005,tannor_nonMarkovian_1999}.
In such dynamics the propagator has the form of Eqn. (\ref{eq:homogeneousEvolution}).

We note here that the above analysis cannot include the influence of the field on the environment, 
since the L-GKS open system dynamics does not include such a mechanism. 

When a weak-field phase-only control is encountered, 
we have to examine how this effect scales with the field coupling strength.
According to the above analysis, while the total population transfer is the leading order in the perturbation, 
i.e., second order in the field coupling strength, 
the phase effect on the population transfer should be the next order, 
i.e., fourth order in the field coupling strength.

In the following section we examine such an example 
and show that the order of the effects are as expected.

\section{Illustrative Example: Population transfer in a four-levels system}

A numerical evaluation of L-GKS open system dynamics which obey the four conditions given in section \ref{sec:4axioms} was performed.
The aim was to examine a case of WFPO control, and check the scaling of the population transfer and phase-dependent phenomena with the field coupling strength.

\subsection{Simulation details}
\begin{figure}[htbp]
\begin{center}
\includegraphics[scale=0.2]{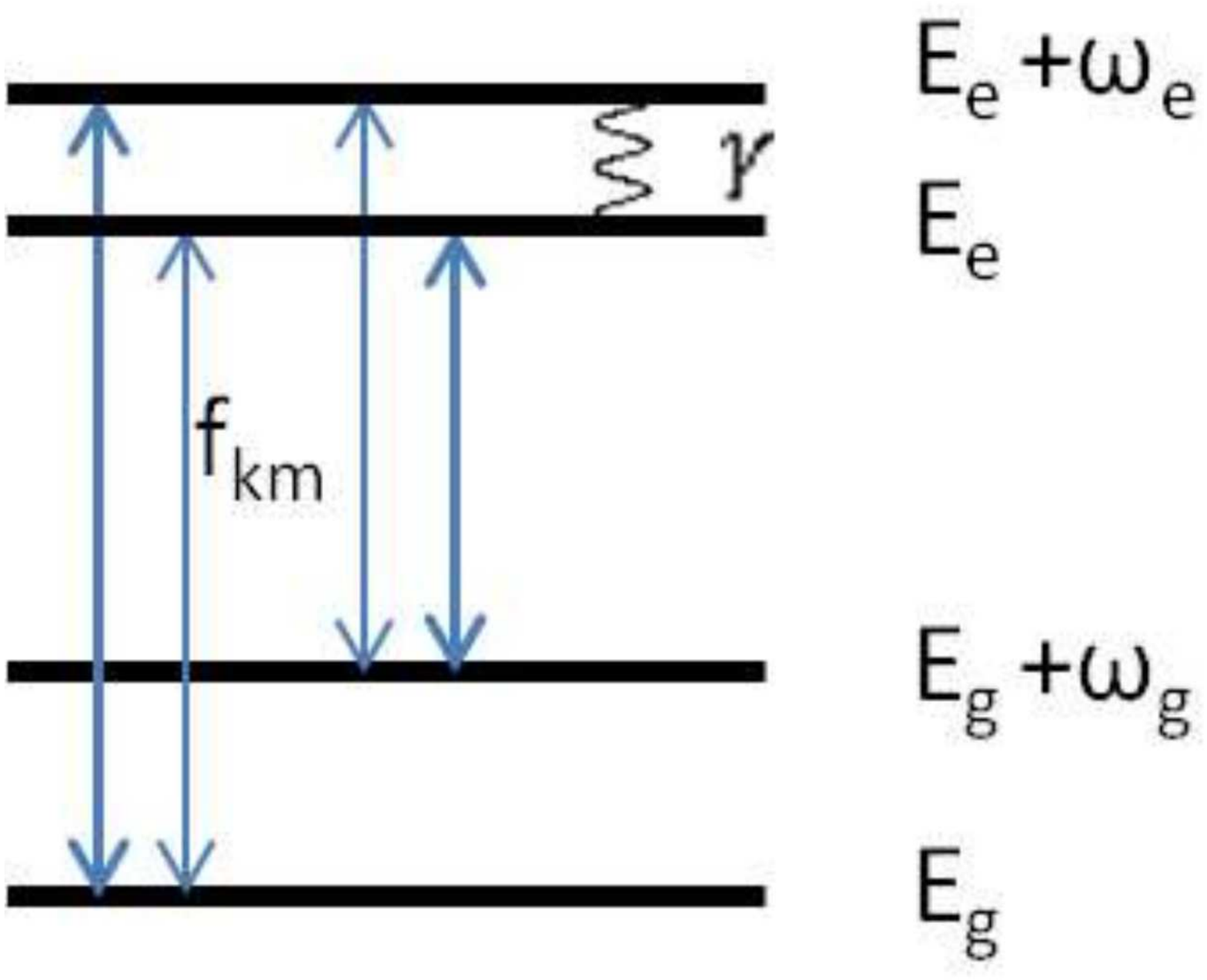} 
\caption{
A schematic diagram of the simulated system:
$E_e$ and $E_g$ are the energies of the surfaces. 
$\omega_e$ and $\omega_g$ are the vibrational frequencies inside the surfaces. 
$f_{km}$ are the Franck-Condon coefficients. 
$\gamma$ is the relaxation coefficient. 
}
\label{fig:sys_diag}
\end{center}
\end{figure}

The system under study is driven by a chirped gaussian field, 
and coupled to an environment with a L-GKS dissipation.
The system is designed such that the final population transfer is affected by the phase of the external field, namely the chirp.
The coupling to the environment induces relaxation which amplifies the chirp effect.
The details of the simulations follow.
Figure \ref{fig:sys_diag} shows a schematic diagram of the simulated system.

The system has four energy levels: 
Two ground energy levels and two excited ones.
The ground levels serve as the ground electronic surface. 
The two excited levels serve as the excited electronic surface. 
These two levels are coupled to each other by a Lindblad-type dissipator. 
Only the external field couples between the surfaces, 
and the field-free Hamiltonian does not couple between them.
The field-free Hamiltonian is:
\begin{equation}
\label{eq:H0}
\op{H}{_0} = 
\left( 
\begin{array}{cccc} 
	E_e+\omega_e & 0   & 0            & 0 \\ 
	0            & E_e & 0            & 0 \\ 
	0            & 0   & E_g+\omega_g & 0 \\  
	0            & 0   & 0            & E_g \\  
\end{array}
\right),
\end{equation}

where $E_e$ and $E_g$ are the energies of the surfaces, 
while $\omega_e$ and $\omega_g$ are the vibrational frequencies inside the surfaces.
We used the rotating frame for the actual simulations.
Therefore the relevant parameter is the detuning, 
defined by $\delta \equiv E_e-E_g-\omega_L$,
where $\omega_L$ is the carrier frequency (see below).

The ground and excited surfaces are coupled with the field operator:
\begin{equation}
\label{eq:Vt}
\mu \op{V}{(t)} = 
\mu \left( 
\begin{array}{cccc} 
	0 & 0 & f_{24} \varepsilon(t) & f_{14} \varepsilon(t) \\  
	0 & 0 & f_{23} \varepsilon(t) & f_{13} \varepsilon(t) \\  
	f_{24}^* \varepsilon^*(t) & f_{23}^* \varepsilon^*(t) & 0 & 0 \\  
	f_{14}^* \varepsilon^*(t) & f_{13}^* \varepsilon^*(t) & 0 & 0 \\  
\end{array}
\right),
\end{equation}

where $\mu$ is the field coupling strength,
$f_{km}$ are the Franck-Condon coefficients, 
and $\varepsilon(t)$ is the external field applied to the system.
We set the Franck-Condon coefficients to mimic the case of two displaced harmonic oscillators:
$f_{14}$ $f_{23}$ are large, while $f_{24}$ $f_{13}$ are smaller.

\label{sec:the_field}
The goal of these simulations is to examine the dependence of final population transfer on phase properties of the field.
The field we use is a chirped Gaussian pulse. 
We define the chirp at the frequency domain in such a way that changing the chirp changes the phase properties of the field but not the amplitude, as defined above in section \ref{sec:intro}, Eq. (\ref{eq:amp_and_phase}).

\begin{equation}
\tilde{\epsilon}(\omega)
= \frac{1}{\pi^{\frac{1}{4}}\sqrt{\Delta\omega}}
\exp\left( 
- \frac{1}{2} \left(\frac{\omega-\omega_L}{\Delta\omega}\right)^{2}
+ i\chi \left( \omega-\omega_L \right)^{2}
\right),
\end{equation}

with $\Delta\omega$ as the bandwidth, $\chi$ as the chirp, 
and $\omega_L$ is the carrier frequency.

Introducing $\sqrt{\Delta\omega}$ in the pre-exponential factor keeps the total energy of the pulse unchanged while changing the bandwidth, such that:

\begin{equation}
\int_{-\infty}^{\infty}\left|\tilde{\epsilon}(\omega)\right|^{2}\, d\omega=1.
\end{equation}

The inverse FT of the chirped pulse is:

\begin{equation}
\label{eq:field}
\epsilon(t)
= \frac{1}{\pi^{\frac{1}{4}}\sqrt{\tau_{0}-\frac{2i\chi}{\tau_{0}}}}
\exp\left(
-\left(\frac{1}{2}+i\frac{\chi}{\tau_{0}^{2}}\right)
\left(\frac{t}{\tau_{ch}}\right)^{2}
\right)
e^{-i \omega_L t},
\end{equation}

with $ \tau_{0} = \frac{1}{\Delta \omega}$ as the duration of the unchirped pulse, 
and $\tau_{ch} = \omega_{ch}\tau_{0}$ as the extended pulse duration, caused by the chirp:
$\omega_{ch}=\sqrt{1+4\frac{\chi^{2}}{\tau_{0}^{4}}}$.

The environment coupling induces a relaxation from the fourth  energy level to the third one.
The relaxation is described by a L-GKS dissipator, 
which is induced by an annihilation operator $\op{s}{_{34}}=\ket{3}\bra{4}$.
This operator has all-zeros entries, except one entry, which transfers population from the fourth level to the third.

This operator induces coupling inside the excited surface, but not between the surfaces.
The dissipator is: 

\begin{equation}
\supop{L}_D \left[\op{\rho}{}\right] =  
\op{s}{_{34}} \op{\rho}{} \op{s}{_{34}^\dagger} 
-\frac{1}{2}\left( \op{s}{_{34}^\dagger} \op{s}{_{34}} \op{\rho}{} + \op{\rho}{} \op{s}{_{34}^\dagger} \op{s}{_{34}} \right).
\end{equation}

\subsubsection{The dynamics: Equation of motion, initial state and control target}

The equation of motion is:
\begin{equation}
\label{eq:dynamics_equation}
i \hbar \frac{\partial \op{\rho}{}}{\partial t} 
= \supop{L} \op{\rho}{} 
= \com{\op{H}{_0} + \mu \op{V}{(t)},\op{\rho}{}} 
+ i \lambda \supop{L}_D \left[\op{\rho}{}\right].
\end{equation}

Initially, the system is at ground state, i.e., the entire population is on the first level.

Two control targets can be defined and examined:
\begin{itemize}
\item
The final population on the excited surface, i.e. the sum of populations on the third and fourth levels.
In weak fields we expect it to be the leading order in the perturbation strength $\mu$. 
The chirp effect is expected to be in the next order in the perturbation.
\item
The final population on the second level.
The population transfer to this level is in essence a second order process.
The structure of the system makes this population sensitive to chirp sign, promoting cases when higher frequencies precede lower ones (i.e. negative chirps).
In addition, the magnitudes of the Franck-Condon coefficients 
(large $f_{14}$ $f_{23}$, small $f_{24}$ $f_{13}$)
create a scenario where the relaxation in the excited surface enhances the negative-chirp-induced population transfer.
\end{itemize}

The phase-only control effect is examined by performing pairs of simulations in which the only varied parameter is the chirp: 
positive chirp in one simulation and negative in the other.
The difference in the final population on the targets between two simulations in such pairs is defined as the chirp effect.

The values of the parameters used in the simulations are summarized in Table \ref{tab:simu_parmas}.
\begin{table}
\begin{center}
\begin{tabular}{|c|c|c|}
\hline 
Parameter & Value & Unit \\ 
\hline 
$\omega_g$ & 0.5 & [time]$^{-1}$ \\ 
\hline 
$\omega_e$ & 0.1 & [time]$^{-1}$ \\ 
\hline 
$\delta$ & 0.2 & [time]$^{-1}$ \\ 
\hline 
$\mu$ & (several) & [time]$^{-1}$ \\ 
\hline 
$\lambda$ & (several) & [time]$^{-1}$ \\ 
\hline 
$f_{14}$, $f_{23}$ & 0.9 & (unitless) \\ 
\hline 
$f_{24}$, $f_{13}$ & 0.1 & (unitless) \\ 
\hline 
$\Delta \omega$ & 1 & [time]$^{-1}$ \\ 
\hline 
$\chi$ & $\pm$80 & [time]$^2$ \\ 
\hline 
\end{tabular} 
\caption{Simulation parameters}
\label{tab:simu_parmas}
\end{center}
\end{table}
The detuning was selected to maximize the final population transfer.

\subsection{Simulations results}
Simulations were performed with the model described in 
Eqs. (\ref{eq:H0}), (\ref{eq:Vt}), (\ref{eq:field}) and (\ref{eq:dynamics_equation}).
The phase-only control effect was examined by comparing similar simulations where the only difference is the chirp sign: positive or negative. 
The difference of the final population transfer between the two cases is defined as the chirp effect.
The results are presented below.

\subsubsection{Simulation dynamics}
\begin{figure}
\begin{center}
\includegraphics[scale=0.6]{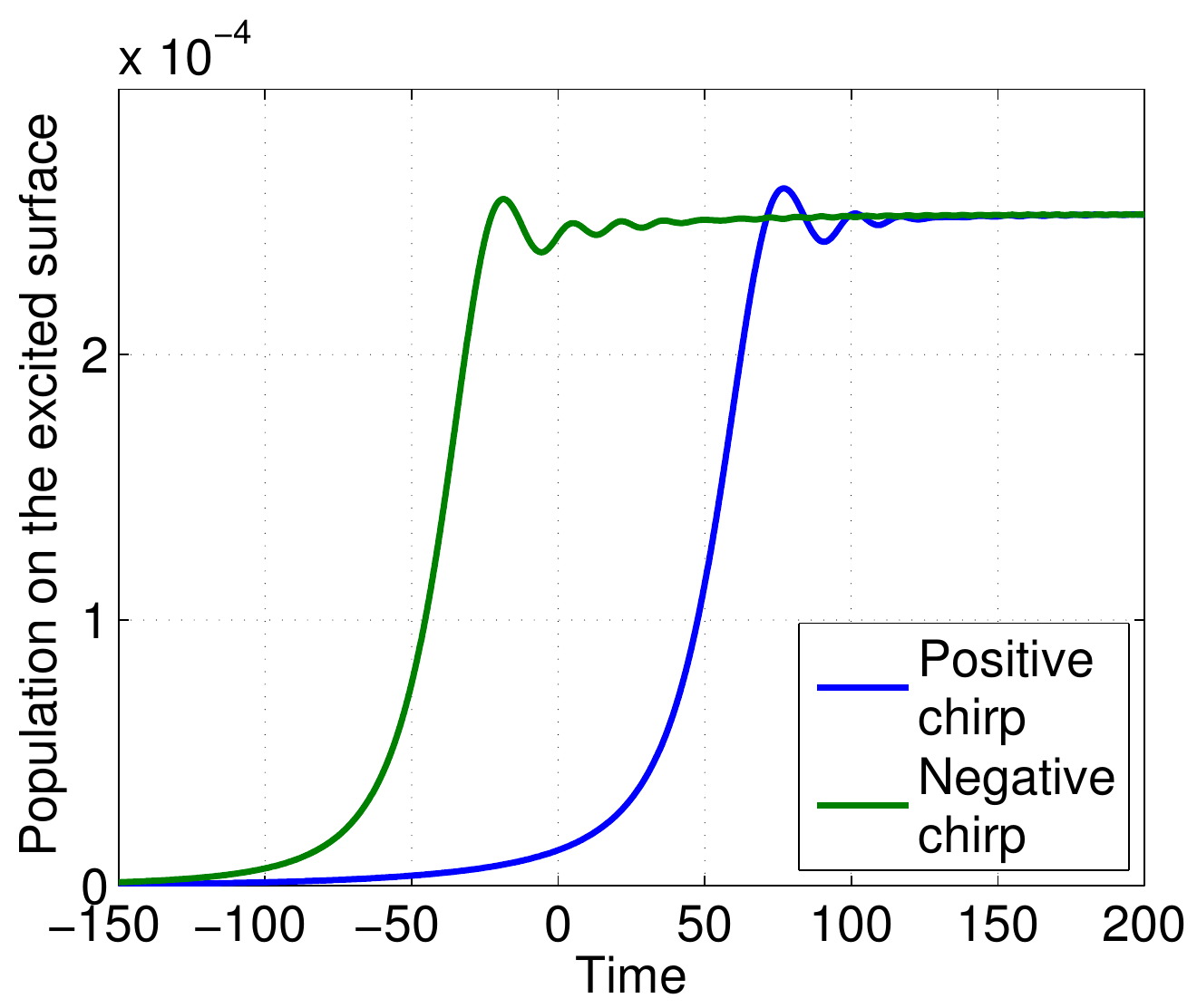} 
\end{center}
\caption{
Population on the excited surface during simulations in which the system is driven 
by positively (blue dashed line) and negatively (green dash-dotted line) chirped fields. 
The population transfer to this surface is a first order process, 
and therefore the difference in the final population, 
which is governed by the next order, cannot be seen on this scale.
}
\label{fig:pt_vs_t_34}
\end{figure}

\begin{figure}
\begin{center}
\includegraphics[scale=0.6]{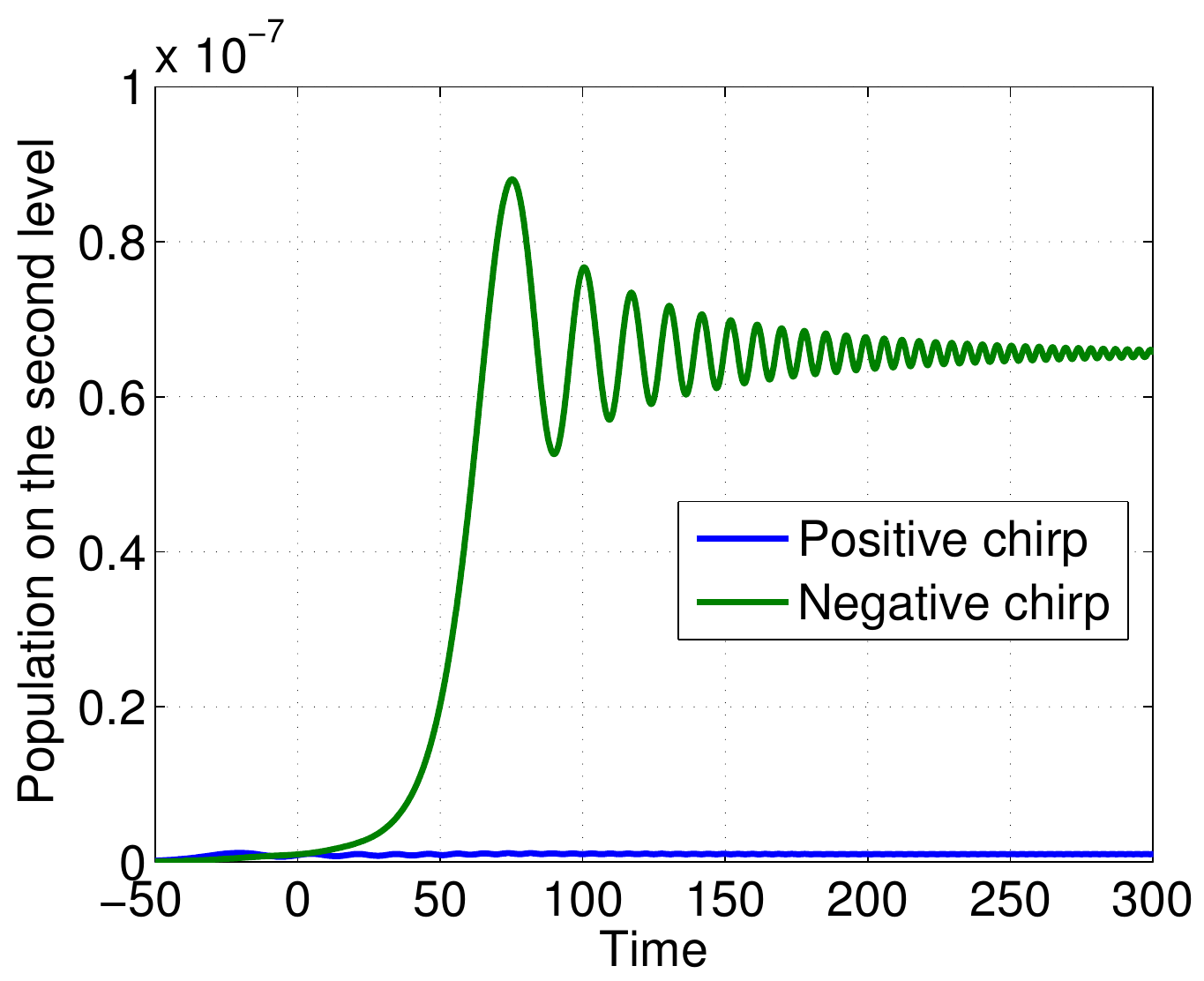} 
\end{center}
\caption{
Population on the second level, during the same simulations as in Figure \ref{fig:pt_vs_t_34}. 
Note that the scale in this figure is different.
This population is a second order process in essence,
and therefore is controlled by the chirp: 
Positive chirp yields a very small population transfer to the second level, 
while negative chirp yields population transfer which is two order of magnitudes larger (although still small).
}
\label{fig:pt_vs_t_2}
\end{figure}

Figure \ref{fig:pt_vs_t_34} shows an example of the population of the exited surface during the simulations of the positive and negative chirp.
The population transfer to this surface is a first order process, 
and therefore the difference in the final population, which is governed by the next order, cannot be seen on this scale.
The population of the second level is presented in Figure \ref{fig:pt_vs_t_2}. 
This population is a second order process in essence (note the different scale),
and therefore is controlled by the chirp: 
Positive chirp yields a very small population transfer to the second level, 
while negative chirp yields population transfer which is by two orders of magnitudes larger.

\subsubsection{Relaxation-induced chirp effect}
\begin{figure}
\begin{center}
\includegraphics[scale=0.7]{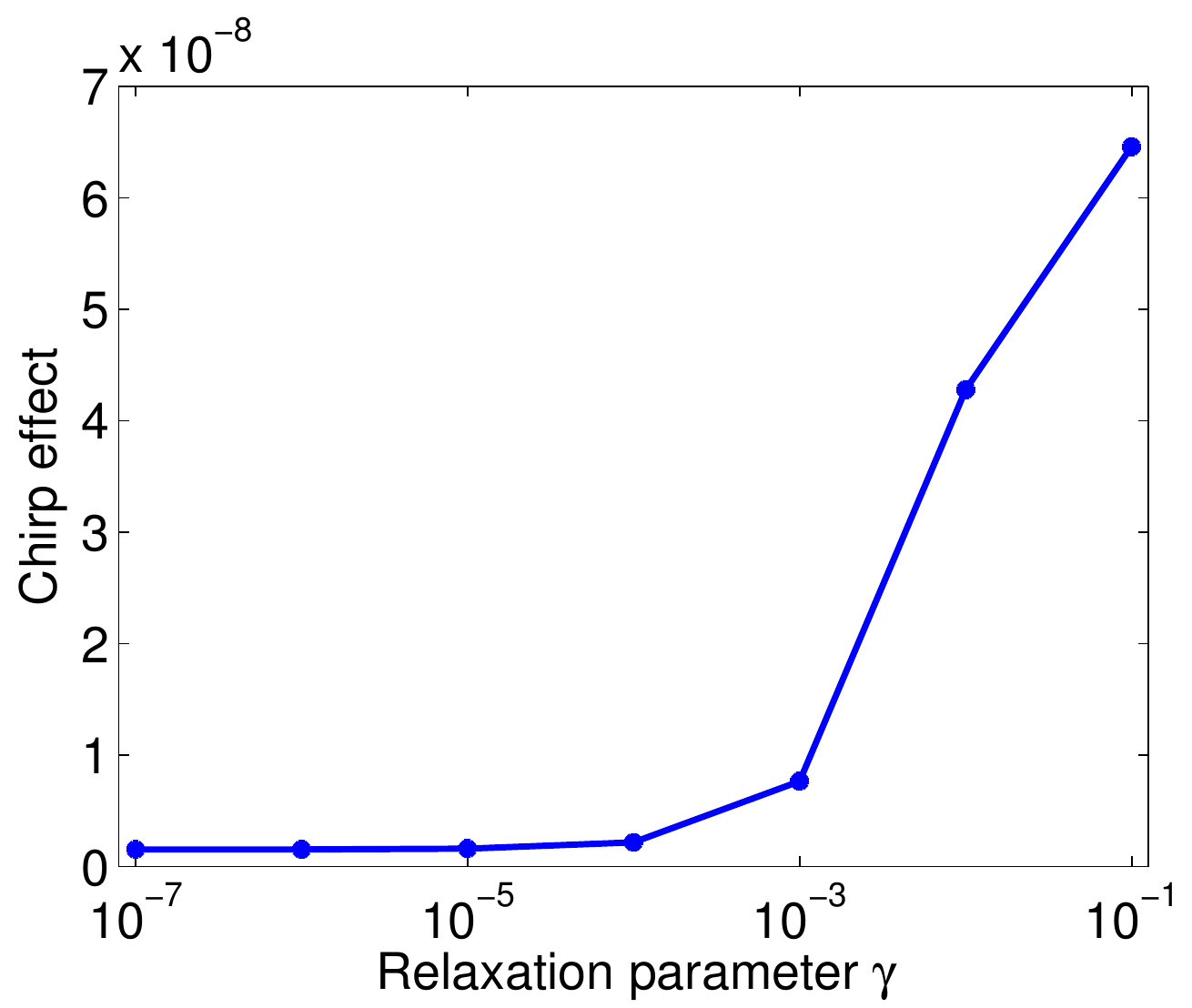} 
\end{center}
\caption{
Chirp effect versus the relaxation coefficient $\gamma$. 
The chirp effect is defined as the difference of the final population transfer between the simulation with positive chirp and the simulation with the negative chirp.
X-axis is log-scale.
The chirp effect is enhanced by the relaxation process.
}
\label{fig:chiEff_vs_g}
\end{figure}

Figure \ref{fig:chiEff_vs_g} presents the chirp effect as a function of the relaxation coupling coefficient $\gamma$.
The chirp effect is enhanced by the relaxation process. 
In the following, we will show that despite that enhancement, 
the chirp effect still scales as the fourth order of the field strength.

\subsubsection{The scaling of the population transfer and the chirp effect with the field strength}

\begin{figure}
\begin{center}
\includegraphics[scale=0.7]{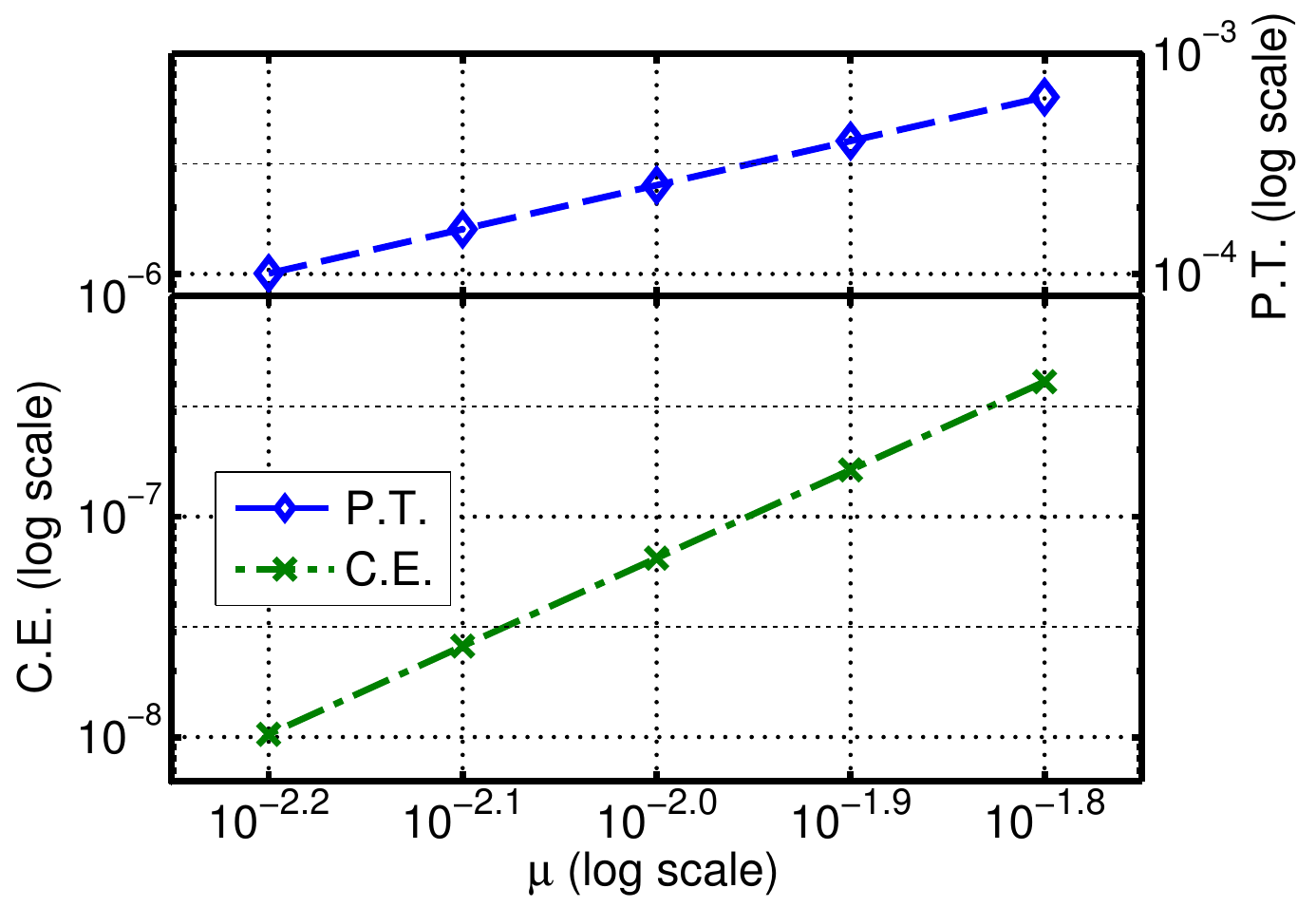} 
\end{center}
\caption{
The final population transfer (P.T., upper panel) to the upper surface 
and the chirp effect (C.E, lower panel) of this P.T.
vs. the field strength $\mu$, on log-log scale. 
Note that there is a gap in the Y-axis 
(emphasized by the labels on the right side in the upper panel), 
although both lines are on the same scale.
The slope of the P.T. is 2, i.e. the P.T. scales as $\mu^2$.
The slope of the C.H. is 4, i.e. the C.E. scales as $\mu^4$.
}
\label{fig:pt_excited_vs_m}
\end{figure}

\begin{figure}
\begin{center}
\includegraphics[scale=0.7]{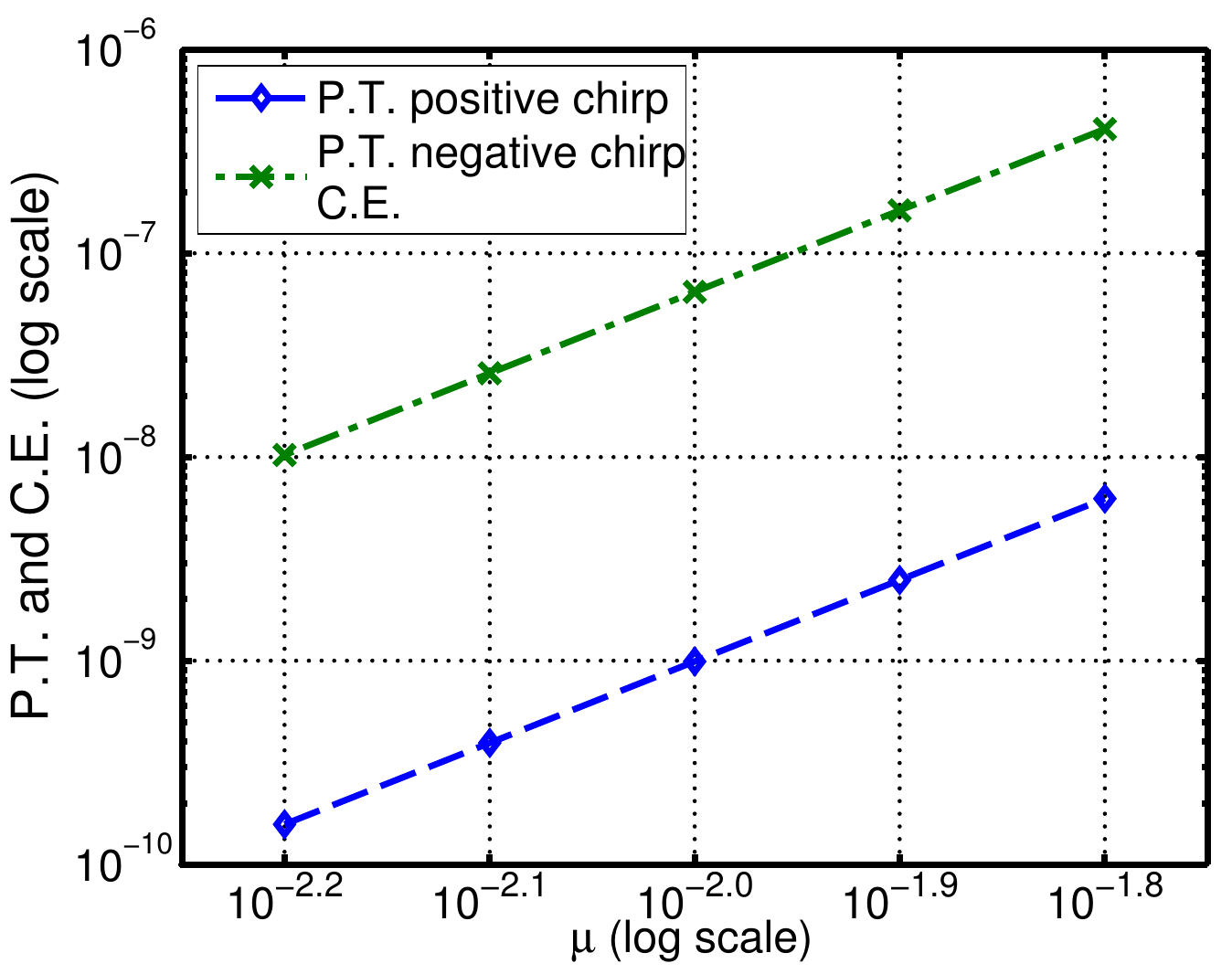} 
\end{center}
\caption{
The final population transfer (P.T.) to the second level:
The lower line shows the P.T. for positive chirps.
The upper line shows the P.T. for negative chirps, 
which almost equals the chirp effect (C.E.). 
Both lines are plotted vs. the field strength $\mu$, on log-log scale. 
The slope of the both lines is 4, i.e. they scale as $\mu^4$.
}
\label{fig:pt_second_vs_m}
\end{figure}

We examined the scaling of the population transfer and the chirp effects with the strength of the external field.

Figure \ref{fig:pt_excited_vs_m} shows the results for the target on the excited state.
As expected, 
we found that the slope of the population transfer is 2, 
i.e. the population transfer scales as $\mu^2$,
while the slope of the chirp effect is 4, 
i.e. the chirp effect scales as $\mu^4$.

Figure \ref{fig:pt_second_vs_m} shows the results for the target on the second level.
Essentially, the population transfer to this level is of the next order, which is in the same order of the chirp effect.
Therefore, we expect to find the same scaling with field strength for both phenomena.
Actually, the population transfer to this level in the case of positive chirps is very small, and almost vanishes,
and therefore the chirp effect and the population transfer for negative chirp are almost the same.
As expected, 
we found that the slope of population transfer for both chirp signs, 
as well as the slope of the chirp effect is four, 
i.e. they all scale with field strength as $\mu^4$.

\section{Conclusions}

The issue of the weak field phase only control is of fundamental importance. Molecular spectroscopy in condensed phase
assumes that the energy absorbed for each frequency component  in the linear regime depends only on the molecular properties.
At normal temperatures the molecule is in its ground electronic surface. By relating the energy absorbed to the population transfer
we find that the validity of molecular spectroscopy in
condensed phase relies on the impossibility of WFPO. 
Brumer and co-workers have studied extensively this phenomena
\cite{spanner_one-photon-control_2010, brumer_onePhoton_faraday_2013, brumer_onePhoton_JCP_2013}.
The present study is in line with these findings. For a molecular system modelled by the L-GKS Markovian dynamics WFPO
is impossible for observables which are invariant to the field free dynamics. 

The method of proof, based on functional derivative (cf. Appendix \ref{sec:app_phase_indep}, 
can be extended to other scenarios.

The numerical model is also consistent with the work of
Konar, Lozovoy and Dantus \cite{dantus_chirp_2012} showing fourth order scaling of the chirp effect with the driving field strength.
Contrary to their finding that the positive chirp is sensitive to the solvent  
\cite{dantus_solvent_chirp_2014}, 
our numerical model finds strong sensitivity to negative chirp.

Shapiro and Han \cite{shapiro_LinearResponse_2012} argue that apparent linear response 
experimental phenomena  are not necessarily weak-field effects. 
In the present study, 
the analysis is based on order by order perturbation theory and addresses this issue. 
Experimental or numerical tests have to be extremely careful in checking the scaling order of the effect.

Readdressing the theme of the study: 
Is there a weak field phase only control in open systems?
We obtained a partial answer.
Under Markovian L-GKS dynamics WFPO is impossible. This still leaves open the possibility of WFPO in non-Markovian scenarios.
The main assumption that should be challenged is the tensor product separability of the system and bath in L-GKS dynamics.
Preliminary numerical evidence from non separable system-bath models may point to the possibility of WFPO for population
transfer with enhancement for positive chirp. More work is required to establish this possibility.

\section*{Acknowledgments}
Work supported by the Israel Science Foundation (ISF).
We want to thank Brian Burrows, Gil Katz, Amikam Levy, Ido Schaefer,
Dwayne Miller and Valentyn Prokhorenko
 for useful discussions.

\appendix

\section{Appendix: The phase independence of the autocorrelation function} \label{sec:app_phase_indep}

The autocorrelation function (ACF) of the field $\epsilon(t)$ is the inverse Fourier transform (FT) of the spectral density of the field, 
$J(\omega) = \left| \tilde{\epsilon}(\omega) \right|^2$.
In this paper, the population transfer to second order is proportional to the Laplace transform of the ACF (cf. Eq. \ref{eq:integral_unitary}).
Therefore a careful examination of the phase properties in this case are required.
First, we derive the phase independence of the ACF. 
Similarly, the phase independence of the cross-correlation function of the field with its derivative is obtained.
We use the functional derivative of these two correlation functions 
to prove the phase independence of the absorption spectrum. 

The autocorrelation function is defined as:
\begin{equation}
C(t) = \int\limits_{-\infty}^{\infty} \,d\tau
\epsilon(\tau+t)\epsilon^*(\tau) .
\end{equation}

Similarly, the cross-correlation function of the field with its derivative is defined as:
\begin{equation}
D(t) = \int\limits_{-\infty}^{\infty} \,d\tau
\left.\frac{\partial\epsilon}{\partial t}\right|_{\tau+t}
\epsilon^*(\tau) 
\end{equation}

We will use the spectral representation of the field:
\begin{equation}
\epsilon(t) 
= \int\limits_{-\infty}^{\infty} \,d\omega \tilde{\epsilon}(\omega) e^{-i\omega t}
= \int\limits_{-\infty}^{\infty} \,d\omega 
\tilde{A}(\omega) e^{i \tilde{\varphi}(\omega)} e^{-i\omega t},
\end{equation}
where the real functions $ \tilde{A}(\omega) $ and $ \tilde{\varphi}(\omega) $ 
are the amplitude and phase, respectively.
The spectral representation of the field derivative equals the spectral representation of the field, 
multiplied by $(-i \omega)$:
\begin{equation}
\frac{\partial\epsilon(t)}{\partial t}=\frac{\partial}{\partial t}\int\limits _{-\infty}^{\infty}\, d\omega\tilde{\epsilon}(\omega)e^{-i\omega t}
= \int\limits _{-\infty}^{\infty}\, d\omega (-i \omega)\tilde{\epsilon}(\omega)e^{-i\omega t}
\end{equation}

The functional derivatives of these correlation functions with respect to the phase are:
\begin{equation}
\frac{\delta C(t)}{\delta \tilde{\varphi}(\omega)}
= 
\int\limits_{-\infty}^{\infty} \,d\tau
\left\lbrace
\frac{\delta \epsilon(\tau+t)}{\delta \tilde{\varphi}(\omega)} \epsilon^*(\tau)
+\epsilon(\tau+t) \frac{\delta \epsilon^*(\tau) }{\delta \tilde{\varphi}(\omega)}
\right\rbrace
\end{equation}

\begin{equation}
\frac{\delta D(t)}{\delta\tilde{\varphi}(\omega)}=\int\limits _{-\infty}^{\infty}\, d\tau\left\lbrace \frac{\delta\left(\left.\frac{\partial\epsilon}{\partial\tau}\right|_{\tau+t}\right)}{\delta\tilde{\varphi}(\omega)}\epsilon^{*}(\tau)+\left.\frac{\partial\epsilon}{\partial\tau}\right|_{\tau+t}\frac{\delta\epsilon^{*}(\tau)}{\delta\tilde{\varphi}(\omega)}\right\rbrace 
\end{equation}

We need the following functional derivatives with respect to the phase:
\begin{equation}
\begin{array}{c}
\frac{\delta \epsilon(t)}{\delta \tilde{\varphi}(\omega)}
= i \tilde{A}(\omega) e^{i \tilde{\varphi}(\omega)} e^{-i\omega t}
= i \tilde{\epsilon}(\omega) e^{-i\omega t} \\
\frac{\delta \epsilon^*(t)}{\delta \tilde{\varphi}(\omega)}
= -i \tilde{A}(\omega) e^{-i \tilde{\varphi}(\omega)} e^{i\omega t}
= -i \tilde{\epsilon}^*(\omega) e^{i\omega t} \\
\frac{\delta\left(\frac{\partial\epsilon(t)}{\partial t}\right)}{\delta\tilde{\varphi}(\omega)}
=i(-i\omega)\tilde{A}(\omega)e^{i\tilde{\varphi}(\omega)}e^{-i\omega t}
=i(-i\omega)\tilde{\epsilon}(\omega)e^{-i\omega t} .\\
\end{array}
\end{equation}

Substituting in the functional derivative of the correlation functions, 
we get (changing integration variable in the second line $\tilde{\tau}=\tau+t$): 

\begin{equation}
\begin{array}{rcl}
\frac{\delta C(t)}{\delta\tilde{\varphi}(\omega)} & = &
\int\limits_{-\infty}^{\infty} \,d\tau
\left[
i \tilde{\epsilon}(\omega) e^{-i\omega (\tau+t)} \epsilon^*(\tau)
- i \epsilon(\tau+t) \tilde{\epsilon}^*(\omega) e^{i\omega \tau}
\right]
\\
& = &
i \tilde{\epsilon}(\omega) e^{-i\omega t}
\left[
\int\limits_{-\infty}^{\infty} \,d\tau e^{i\omega \tau} \epsilon(\tau)
\right] ^*
- i \left[
\int\limits_{-\infty}^{\infty} \,d\tilde{\tau}
\epsilon(\tilde{\tau}) e^{i\omega (\tilde{\tau}-t)}
\right]
\tilde{\epsilon}^*(\omega) 
\\
& = &
i \tilde{\epsilon}(\omega) e^{-i\omega t}
\tilde{\epsilon}^*(\omega) 
- i \tilde{\epsilon}(\omega) e^{-i\omega t}
\tilde{\epsilon}^*(\omega) 
\\
& = & 0 ,\\
\end{array}
\end{equation}

and, similarly,
\begin{equation}
\begin{array}{rcl}
\frac{\delta D(t)}{\delta\tilde{\varphi}(\omega)} & = &
\int\limits_{-\infty}^{\infty} \,d\tau
\left[
i(-i\omega)\tilde{\epsilon}(\omega)e^{-i\omega(\tau+t)}\epsilon^{*}(\tau)
-i\left.\frac{\partial\epsilon}{\partial t}\right|_{\tau+t}
\tilde{\epsilon}^{*}(\omega)e^{i\omega\tau}
\right]
\\
& = &
i (-i\omega) \tilde{\epsilon}(\omega) e^{-i\omega t}
\left[
\int\limits_{-\infty}^{\infty} \,d\tau e^{i\omega \tau} \epsilon(\tau)
\right] ^*
- i \left[
\int\limits_{-\infty}^{\infty} \,d\tilde{\tau}
\left.\frac{\partial\epsilon}{\partial t}\right|_{\tilde{\tau}} e^{i\omega (\tilde{\tau}-t)}
\right]
\tilde{\epsilon}^*(\omega) 
\\
& = &
i (-i\omega) \tilde{\epsilon}(\omega) e^{-i\omega t}
\tilde{\epsilon}^*(\omega) 
- i (-i\omega) \tilde{\epsilon}(\omega) e^{-i\omega t}
\tilde{\epsilon}^*(\omega) 
\\
& = & 0 .\\
\end{array}
\end{equation}

\section{Appendix: Detailed calculation of the population transfer}
\label{sec:app_calculations}
We show here the details of the calculations. 

\subsection{Unitary dynamics generated by the Hamiltonian}
Fron Eq. (\ref{eq:commutator_operation}) we get (the operator $\op{V}{}$ transfers population between the surfaces):
\begin{equation}
\supop{V}(t_1) \ket{a}\bra{a} = 
\sum_{\substack {a\in g.s. \\ b\in e.s }} P(a)
 \left(\epsilon(t_1)\mu_{ba}\ket{b}\bra{a} - \epsilon^*(t_1)\mu_{ab}\ket{a}\bra{b}\right).
\end{equation}

Next, we operate with the propagator $ e^{-\frac{i}{\hbar} \supop{L}_0 (t_2-t_1)} $.
When the dynamics is unitary, the Lindbladian includes only the commutator with the Hamiltonian, 
and the  propagation of an element in the density matrix \ket{c}\bra{d} is simply a multiplication by $e^{-i \omega_{cd}t}$,
where $ \omega_{cd}\equiv\omega_{c}-\omega_{d} $, so we get:
\begin{equation}
\sum_{\substack {a\in g.s. \\ b\in e.s }} P(a) 
\left(\epsilon(t_1)\mu_{ba}\ket{b}\bra{a} e^{-i \omega_{ba}(t_2-t_1)} 
- \epsilon^*(t_1)\mu_{ab}\ket{a}\bra{b}e^{-i \omega_{ab}(t_2-t_1)}\right).
\end{equation}

Now, we operate with  $\supop{V}(t_2)$, to get (using Eq. (\ref{eq:commutator_operation})):
\begin{equation}
\sum\limits_{\substack {a\in g.s. \\ b\in e.s \\k }} P(a) 
\left\lbrace \epsilon(t_1)\epsilon^*(t_2)\mu_{ba} 
		\left( \mu_{kb}\ket{k}\bra{a} - \mu_{ak}\ket{b}\bra{k} \right) e^{-i \omega_{ba}(t_2-t_1)} + h.c. \right\rbrace,
\end{equation}
where $h.c.$ stands for hermitian conjugate.

Now we project on the excited surface (with \op{P}{_e}), and perform the trace.
For a general element in the density matrix \ket{c}\bra{d}, 
we do so by taking the sum of diagonal matrix elements that belong to the excited surface:
$
\sum\limits_{m\in e.s.} \braket{m}{c}\braket{d}{m}, 
$
\ so we get
\begin{equation}
- \sum\limits_{\substack {a\in g.s. \\ b\in e.s \\k \\m\in e.s.}} P(a)
\left\lbrace \epsilon(t_1)\epsilon^*(t_2)\mu_{ba} \mu_{ak}\braket{m}{b}\braket{k}{m} 
e^{-i \omega_{ba}(t_2-t_1)} + c.c. \right\rbrace.
\end{equation}

\braket{m}{b} and \braket{k}{m} are $\delta_{mb}$ and $\delta_{km}$, respectively.
When we sum over $k$ and $m$ we get
\begin{equation}
- \sum\limits_{\substack {a\in g.s. \\ b\in e.s}} P(a)
\left\lbrace \epsilon(t_1)\epsilon^*(t_2)\left| \mu_{ab} \right|^2 e^{-i \omega_{ba}(t_2-t_1)} + c.c. \right\rbrace.
\end{equation}

Next, we integrate over $t_1$ and $t_2$. 
Since the pulse has a finite duration, we can extend the integration limits to  $(-\infty,\infty)$:
\begin{equation}
\sum\limits_{\substack {a\in g.s. \\ b\in e.s}} P(a) 
\frac{\left| \mu_{ab} \right|^2}{\hbar^2}
\int\limits_{-\infty}^{\infty} \,dt_2 \int\limits_{-\infty}^{t_2} \,dt_1
\left\lbrace \epsilon(t_1)\epsilon^*(t_2) e^{-i \omega_{ba}(t_2-t_1)} + c.c. \right\rbrace.
\end{equation}

We change variables in the integral, from $t_2$ to $ \tau = t_2 - t_1$, and we get the integral:
\begin{equation}
\int\limits_{0}^{\infty} \,d\tau 
\left( \int\limits_{-\infty}^{\infty} \,dt_1
\epsilon(t_1)\epsilon^*(\tau+t_1) \right) 
e^{-i \omega_{ba}\tau} 
= \int\limits_{0}^{\infty} \,d\tau 
C^*(\tau)
e^{-i \omega_{ba}\tau} ,
\end{equation}
where $C(\tau)$ is the complex conjugate of the autocorrelation function 
(ACF) of the field $\epsilon(t)$ (defined above in Appendix \ref{sec:app_phase_indep}).
Finally, we have:
\begin{equation} 
\Delta N = \sum\limits_{\substack {a\in g.s. \\ b\in e.s}} P(a) 
\frac{\left| \mu_{ab} \right|^2}{\hbar^2}
\left\lbrace \int\limits_{0}^{\infty} \,d\tau 
C^*(\tau) e^{-i \omega_{ba}\tau} + c.c. \right\rbrace.
\end{equation}

We see that the population transfer does not depend directly on the field, 
only through the field's ACF.
This result is not new \cite{spanner_one-photon-control_2010}.
It is presented here in order to demonstrate the perturbative calculation in Liouville space,
and to emphasise the dependence on the ACF.

\subsection{General Lindbladian-generated dynamics}

We show here that the population transfer depends on the field only through the ACF 
also in QDS description of non unitary dynamics.

Consider a Linbladian that can induce dephasing and relaxation inside the electronic surfaces, but not between them.
We do not treat here electronic dephasing or electronic relaxation.
Population transfer is done only by \supop{V} (the commutator of \op{V}{}).

Here, we use a more formal notation: 
we do not write explicitly the matrix elements of the operator \op{\mu}{}.
Instead, for a state \ket{a} (or \bra{a}) in the ground surface, 
we write $\ket{\theta_a}\equiv \op{\mu}{}\ket{a}$  (or $\bra{\theta_a}\equiv \bra{a}\op{\mu}{}$, respectively), 
and it should be understood as a state in the excited electronic surface. 
Also ,we will use the notation $\op{\Theta}{_a} \equiv \ket{\theta_a}\bra{a}$ for the relevant density matrix element. 
We also do not write explicitly the resulting states of the propagation by $\supop{L}_0$,
and write instead expressions like $ e^{-i \supop{L}_0 t} \op{\Theta}{_a} $.

Starting with Eq. (\ref{eq:pop_trans_expr}), and the initial state of Eq. (\ref{eq:init_rho}),
We first operate with $\supop{V}(t_1)$ to get
\begin{equation}
\sum\limits_{a\in g.s.}P(a) 
\left( \epsilon(t_1)\ket{\theta_a}\bra{a} 
- \epsilon^*(t_1)\ket{a}\bra{\theta_a} \right).
\end{equation}
Since \ket{a} and \bra{a} are in the ground surface,
and since $\ket{\theta_a}$ and $\bra{\theta_a}$ are in the excited surface,
$\ket{\theta_a}\bra{a}=\op{\Theta}{_a}$ and $\ket{a}\bra{\theta_a}=\op{\Theta}{_a}^{\dagger}$ are off-diagonal blocks in the density matrix.

Next, we operate with the propagator $ e^{-\frac{i}{\hbar} \supop{L}_0 (t_2-t_1)} $ to get:
\begin{equation}
\sum\limits_{a\in g.s.}P(a) 
\left( \epsilon(t_1) e^{-\frac{i}{\hbar} \supop{L}_0 (t_2-t_1)} \op{\Theta}{_a} 
- \epsilon^*(t_1) e^{-\frac{i}{\hbar} \supop{L}_0 (t_2-t_1)} \op{\Theta}{_a}^\dagger \right).
\end{equation}
Again, the two terms here are off diagonal blocks.

When we operate with $\supop{V}(t_2)$ we get four terms. 
Two of them belong to the ground surface, 
and therefore will be omitted in the projection on the excited surface.
The other terms are:
\begin{equation}
\sum\limits_{a\in g.s.}P(a) 
\left( \epsilon(t_1)\epsilon^*(t_2) \left[ e^{-\frac{i}{\hbar} \supop{L}_0 (t_2-t_1)} \op{\Theta}{_a}\right]\op{\mu}{} 
+c.c. \right)
\end{equation}

Finally, like the previous calculations, 
we perform the trace, extend the integration limits, 
change one integration variable and integrate over the other variable, 
to get the autocorrelation of the field:
\begin{equation}
\Delta N = \frac{1}{\hbar ^2}
\sum\limits_{\substack {a\in g.s. \\ b\in e.s}} P(a) 
\int\limits_{0}^{\infty} \,d\tau \left( C^*(\tau)
\bra{b}\left[ e^{-\frac{i}{\hbar} \supop{L}_0 \tau} \op{\Theta}{_a}\right]\ket{\theta_b} 
+c.c. \right)
\end{equation}
We have to obtain more details on the operation of \op{\mu}{} and $\supop{L}_0$ 
in order to evaluate this expression further, 
but we see that also here the dependence on the field is only through its ACF.

\bibliography{weak_field1} 

\begin{thebibliography}{10}

\bibitem{rice1992new}
Stuart~A Rice.
\newblock New ideas for guiding the evolution of a quantum system.
\newblock {\em Science}, 258(5081):412--413, 1992.

\bibitem{brumer_one-photon-control_1989}
Paul Brumer and Moshe Shapiro.
\newblock One photon mode selective control of reactions by rapid or shaped
  laser pulses: An emperor without clothes?
\newblock {\em Chemical Physics}, 139(1):221--228, December 1989.

\bibitem{spanner_one-photon-control_2010}
Michael Spanner, Carlos~A Arango, and Paul Brumer.
\newblock Communication: Conditions for one-photon coherent phase control in
  isolated and open quantum systems.
\newblock {\em Journal of Chemical Physics}, 133(15):151101, October 2010.

\bibitem{prokhorenko_population-transfer_2005}
Valentyn~I. Prokhorenko, Andrea~M. Nagy, and R.~J.~Dwayne Miller.
\newblock Coherent control of the population transfer in complex solvated
  molecules at weak excitation. an experimental study.
\newblock {\em Journal of Chemical Physics}, 122(18):184502--184502--11, May
  2005.

\bibitem{prokhorenko_retinal-control_2006}
Valentyn~I. Prokhorenko, Andrea~M. Nagy, Stephen~A. Waschuk, Leonid~S. Brown,
  Robert~R. Birge, and R.~J.~Dwayne Miller.
\newblock Coherent control of retinal isomerization in bacteriorhodopsin.
\newblock {\em Science}, 313(5791):1257 --1261, 2006.

\bibitem{vanderwalle_solvation-induced_2009}
P.~van~der Walle, M.~T.~W. Milder, L.~Kuipers, and J.~L. Herek.
\newblock Quantum control experiment reveals solvation-induced decoherence.
\newblock {\em Proceedings of the National Academy of Sciences}, 106(19):7714
  --7717, May 2009.

\bibitem{katz_weak-field-control_2010}
Gil Katz, Mark~A Ratner, and Ronnie Kosloff.
\newblock Control by decoherence: weak field control of an excited state
  objective.
\newblock {\em New Journal of Physics}, 12(1):015003, 2010.

\bibitem{kraus1971}
Karl Kraus.
\newblock General state changes in quantum theory.
\newblock {\em Annals of Physics}, 64(2):311--335, 1971.

\bibitem{lindblad_generators_1976}
G.~Lindblad.
\newblock On the generators of quantum dynamical semigroups.
\newblock {\em Communications in Mathematical Physics}, 48(2):119--130, 1976.

\bibitem{gorini_QDS_1976}
Vittorio Gorini, Andrzej Kossakowski, and E.~C.~G. Sudarshan.
\newblock Completely positive dynamical semigroups of n-level systems.
\newblock {\em Journal of Mathematical Physics}, 17(5):821--825, 1976.

\bibitem{kosloff1992excitation}
Ronnie Kosloff, Audrey~Dell Hammerich, and David Tannor.
\newblock Excitation without demolition: Radiative excitation of ground-surface
  vibration by impulsive stimulated raman scattering with damage control.
\newblock {\em Physical review letters}, 69(15):2172, 1992.

\bibitem{ashkenazi1997quantum}
Guy Ashkenazi, Uri Banin, Allon Bartana, Ronnie Kosloff, and Sanford Ruhman.
\newblock Quantum description of the impulsive photodissociation dynamics of i-
  3 in solution.
\newblock {\em Advances in Chemical Physics, Volume 100}, pages 229--315, 1997.

\bibitem{yan_HEOM_2008}
Jinshuang Jin, Xiao Zheng, and YiJing Yan.
\newblock Exact dynamics of dissipative electronic systems and quantum
  transport: Hierarchical equations of motion approach.
\newblock {\em The Journal of Chemical Physics}, 128(23):--, 2008.

\bibitem{tanimura_HEOM_2005}
Akihito Ishizaki and Yoshitaka Tanimura.
\newblock Quantum dynamics of system strongly coupled to low-temperature
  colored noise bath: Reduced hierarchy equations approach.
\newblock {\em Journal of the Physical Society of Japan}, 74(12):3131--3134,
  2005.

\bibitem{tannor_nonMarkovian_1999}
Christoph Meier and David~J. Tannor.
\newblock Non-markovian evolution of the density operator in the presence of
  strong laser fields.
\newblock {\em The Journal of Chemical Physics}, 111(8), 1999.

\bibitem{brumer_onePhoton_faraday_2013}
Leonardo~A. Pachon, Li~Yu, and Paul Brumer.
\newblock Coherent one-photon phase control in closed and open quantum systems:
  A general master equation approach.
\newblock {\em Faraday Discuss.}, 163:485--495, 2013.

\bibitem{brumer_onePhoton_JCP_2013}
Leonardo~A. Pachón and Paul Brumer.
\newblock Mechanisms in environmentally assisted one-photon phase control.
\newblock {\em The Journal of Chemical Physics}, 139(16):--, 2013.

\bibitem{dantus_chirp_2012}
Arkaprabha Konar, Vadim~V. Lozovoy, and Marcos Dantus.
\newblock Solvation stokes-shift dynamics studied by chirped femtosecond laser
  pulses.
\newblock {\em The Journal of Physical Chemistry Letters}, 3(17):2458--2464,
  2012.

\bibitem{dantus_solvent_chirp_2014}
Arkaprabha Konar, Vadim~V. Lozovoy, and Marcos Dantus.
\newblock Solvent environment revealed by positively chirped pulses.
\newblock {\em The Journal of Physical Chemistry Letters}, 5(5):924--928, 2014.

\bibitem{shapiro_LinearResponse_2012}
Alex~C. Han and Moshe Shapiro.
\newblock Linear response in the strong field domain.
\newblock {\em Phys. Rev. Lett.}, 108:183002, May 2012.

\end{thebibliography}
\bibliographystyle{unsrt} 

\end{document}